\documentclass[aps,prb,twocolumn,groupedaddress,showpacs]{revtex4}
\usepackage{amsfonts}
\usepackage{amssymb}
\usepackage{amsmath}
\usepackage{dcolumn}
\usepackage{graphicx}
\begin{document}

\title{p-wave Pairing in a Two-Component Fermi System with Unequal Population:
Weak Coupling BCS to Strong Coupling BEC Regimes}
%near Feshbach Resonance
\author{Renyuan Liao}
%\affiliation{Department of Physics, Kent State University, Kent, OH 44242}
\author{Florentin Popescu}
%\affiliation{Department of Physics, Kent State University, Kent, OH 44242}
\author{Khandker Quader}
\affiliation{Department of Physics, Kent State University, Kent, OH 44242}

\date{\today}

\begin{abstract}

We study $p$-wave pairing in a two-component Fermi system with unequal
population across weak-coupling BCS to strong-coupling BEC regimes.  We find 
a rich $m_s=0$ spin triplet p-wave superfluid (SF) ground state (GS) structure as a function of population imbalance. Under a phase stability condition, the ``global" energy minimum is given by a multitude of ``mixed'' SF states formed of linear combinations of  $m=\pm1,0$ sub-states of the $\ell=1$ orbital angular momentum state. Except for the ``pure" SF states, ($\ell=1, m=\pm1$), other states exhibit oscillation in energy with
the relative phase between the constituent gap amplitudes. We also find states with ``local" energy minimum that can be stable at higher polarizations, suggesting a quantum phase transition between the "global"  and  ``local" minima phases driven by polarization. We relate the local and global minimum states to Morse and non-Morse critical points. We calculate the variations of the ground state energy with a relative phase angle, coupling and polarization, and discuss possible consequences for experiments. 

\end{abstract}


\maketitle

\section{Introduction} 

Over past several years there has been sustained experimental and theoretical interest in paired fermion ground states with unconventional pairing symmetry. 
%Of these, observation and understanding of various types of p-wave spin triplet condensates %have been topics of intense study. 
Among these are various types of p-wave spin triplet condensates. 
Correlated electron systems, such as, SrRu$_2$O$_4$~\cite{sr} and ferromagnetic 
superconductors~\cite{fmsc} are believed to possess p-wave triplet symmetry. Fermi systems with unequal species population, as in quark matter~\cite{quark}, magnetic field induced organic superconductors~\cite{organic} and in cold fermion systems with unequally populated hyperfine states~\cite{hyperfine}, add a fascinating dimension.
Discovery of s-wave superfluidity in cold atoms~\cite{BCSBEC} subjected to s-wave Feshbach resonance (FR), and observations of p-wave FR~\cite{REG03,TIC04,ZHA04,SCH05} in $^6Li$ and $^{40}K$ had raised the prospect for observing p-wave superfluidity in cold fermi gases. While this may not be as easy, alternate methods~\cite{alternate} and optical 
lattices~\cite{optical,inter-lat} offer encouraging prospects. Past theory work on 
p-wave superfluidity in the BEC-BCS crossover region include those for a single-component Fermi gas~\cite{GUR05,CHE05,BOT05,OHA05}, as well as for the
two-component case~\cite{TIN05} with equal-species population. However, 
%While there have been fervent theoretical efforts \cite{IMPTH,LQ06} on pairing in %asymmetrical  Fermi systems subject to s-wave FR, 
theory work on p-wave pairing for unequal species population has been 
limited~\cite{inter-lat,imp-theory}.

In this paper, we study p-wave superfluidity in a two-component Fermi system 
with {\it unequal} population, across weak-coupling BCS to strong coupling BEC regimes. 
We focus on the case where inter-species pairing interaction is dominant.  While this is interesting to study on general grounds, we are also motivated by work~\cite{inter-lat}
on population imbalanced Fermi mixtures in optical lattices, and 
by the observation~\cite{TIN05,TIC04} that unlike liquid $^3He$,
pairing interaction in cold atoms may be highly anisotropic in
``spin''-space. (``spin'' referring to hyperfine states). For
example, in $^6Li$, when the hyperfine pair $|m_s,m^{\prime}_s>= |1/2,-1/2>$ is at
resonance, the pairs $|1/2,1/2>$ and $|-1/2,-1/2>$ would not be. So, intra-spin
interactions can be neglected. Pairs in p-wave superfluids 
%(orbital angular momentum $\ell=1$)
with unlike ``spin'' components can, however, have different $\ell = 1$ components, namely, $m= \pm 1,0$;  the gap parameters $\Delta_{\ell=1, m}$ are related to the spherical harmonics, $Y_{1,m=\pm1,0}$.

Our study of the population imbalanced Fermi system lead
to several new results and predictions. These should appeal broadly to systems exhibiting p-wave pairing, especially those with population imbalance between species.
We find a rich superfluid (SF) ground state (GS) structure.
%The p-wave state $\Delta_1$(hence $\Delta_{-1}$ by symmetry) give a GS {\it global} %minimum energy. 
%($\Delta_{1m} \equiv \Delta_m$ hereforth). 
Under a condition on relative phase between three pairing amplitudes $\Delta_{m}$'s
($\Delta_{\ell=1,m} \equiv \Delta_m$ hereforth), a multitude of ``mixed''  superfluid states of the form
$a_0\Delta_0 + a_1\Delta_1 + a_{-1}\Delta_{-1}$ are found to be degenerate with $\Delta_{\pm 1}$; these give the {\it global} minimum GS energy. In addition, we find states with {\it local} minimum energy. We provide a geometric representation of the p-wave SF states (Fig. 1).
%The states exhibiting global minimum
%lie on a ``semicircle''  formed by the intersection of
%the surface of the sphere formed by $\Delta_1, \Delta_{-1}, \Delta_0$ with a plane
%defined by $|\Delta_1| +|\Delta_{-1}|$ = const.
Our zero temperature polarization ($P$) vs p-wave coupling phase diagram (Fig. 2)
shows two superfluid phases, comprising of states with global and local minimum
energy respectively, and a region of phase separation. 
Our results suggest the possibility of a quantum phase transition
between the two superfluid phases, driven by polarization. 
We find the $P \ne$ 0 ground state structure to be preserved in the $P \rightarrow$ 0 limit; hence richer than that obtained earlier \cite{TIN05} for $P$ = 0. The energies of the "mixed" states show oscillations with the relative phase angle (Fig.3), that may be interesting to explore experimentally.
We also study the behavior of superfluid ground state energy with coupling and polarization.

The paper is organized as follows. In Sec. II, we present the model used for describing the two-component Fermi system with unequal spin population across the BCS to BEC regimes. We provide details of the finite-temperature imaginary-time Green's function method used for obtaining the coupled p-wave gap and number equations, and the grand canonical potential. In Sec. III, we carry out a free energy analysis of the 
p-wave ground states in a population imbalanced system. We provide details of the derivation of the free energy to quartic order in pairing gap amplitudes, and the expansion coefficients in terms of the underlying coupling and polarization. This section also shows how minimization of the free energy leads to conditions of stability of p-wave superfluid phases and the nature of the states that may give rise to global and local minima of the free energy. A geometric depiction is provided for a clearer understanding of the possible p-wave states, and the dependence of some of the states on a phase angle. To provide a perspective on the stability of the states, we also present here a brief discussion of connection of the stable states to "critical points" in Catastrophe Theory. In Sec. IV, we discuss our self-consistent solutions of coupled gap and number equations for fixed polarization and coupling parameter. We outline how the free energy expansion coefficients and the free energy are obtained, and the scheme by which the p-wave coupling vs polarization phase diagram across BEC-BCS regimes is constructed (Sec IVA). Our study of the variation of the free-energy with phase angle, coupling and polarization is discussed in Sec IVB. We conclude with a summary of the work and discussions in Sec. V.

\section{Model and Development} 

We consider a Fermi system with
%We consider pairing in a two-component Fermi system with 
unequal ``spin''($(\uparrow,\downarrow)\equiv(1/2,-1/2)$) 
population, and with all fermions having the same mass.
Pairing interactions in all three $\ell=1$ 
channels ($m = 0, \pm 1$) are taken to be equal, and
%Following the rationale above,
intra-species interactions ($\uparrow\uparrow$ or $\downarrow\downarrow$) are set to zero. Since we adjust self-consistently the chemical potential with the strength
and sign of the coupling, in our fermion-only model, molecules would
appear naturally as 2-fermion bound states. The $S=1, m_s=0$ triplet pairing Hamiltonian
is then given by:
\begin{eqnarray}\label{Ham}
{\mathcal{H}}&=&\sum_{\mathbf{k}\sigma}\xi_{\mathbf{k}\sigma}c^{\dag}_{\mathbf{k}\sigma}c_{\mathbf{k}\sigma}\nonumber\\
&+&\sum_{\mathbf{k}\mathbf{k}^{\prime}\mathbf{q}}V_{\mathbf{k}\mathbf{k}^{\prime}}
c^{\dag}_{\mathbf{k}\smash{+}\mathbf{q}/2\uparrow}c^{\dag}_{-\mathbf{k}\smash{+}\mathbf{q}/2\downarrow}
c_{-\mathbf{k}^{\prime}\smash{+}\mathbf{q}/2\downarrow}c_{\mathbf{k}^{\prime}\smash{+}\mathbf{q}/2\uparrow}
\end{eqnarray}
where $c_{\mathbf{k}\sigma}$ ($c^{\dag}_{\mathbf{k}\sigma}$) is the
annihilation (creation) operator for a fermion with momentum
$\mathbf{k}$, kinetic energy
$\xi_{\mathbf{k}\sigma}\smash{=}\epsilon_{\mathbf{k}}\smash{-}\mu_{\sigma}$,
and spin $\sigma$; $\mu_{\sigma}$ is the
chemical potential of each component, and $\epsilon_{\mathbf{k}}\smash = \hbar^{2}k^{2}/2m$.
$V_{\mathbf{kk^{\prime}}}$ is the pairing interaction.
 
We consider condensate pairs with zero center-of-mass momentum, ($\mathbf{q}\smash{=}0$). While non-zero $\mathbf{q}$ would be interesting from the perspective of FFLO~\cite{FFLO} states, the problem of ${\bf q} = 0$ unconventional pairing in systems with unequal population is rich in itself. 
${\mathcal{H}}$ is mean-field (MF) decoupled via the
$S=1,m_s=0$ ``spin''-triplet pairing gap function
$\Delta_{\downarrow\uparrow}(\mathbf{k})\equiv
\Delta(\mathbf{k})\smash{=}\smash{-}\sum_{\mathbf{k}^{\prime}}
V_{\mathbf{k}\mathbf{k}^{\prime}}\langle
c_{-\mathbf{k}^{\prime}\downarrow}
c_{\mathbf{k}^{\prime}\uparrow}\rangle$
giving:
\begin{eqnarray}\label{Ham-MF}
{\mathcal{H}}^{\mathrm{MF}}&=&\sum_{\mathbf{k}\sigma}\xi_{\mathbf{k}\sigma}c^{\dag}_{\mathbf{k}\sigma}c_{\mathbf{k}\sigma}\; - \;\;\!\!\sum_{\mathbf{k}}
\left[\Delta(\mathbf{k})c^{\dag}_{\mathbf{k}\uparrow}
c^{\dag}_{-\mathbf{k}\downarrow}\smash{+}\mathrm{H.c.}\right]\nonumber\\
%&-&\!\!\sum_{\mathbf{k}}
%\left[\Delta(\mathbf{k})c^{\dag}_{\mathbf{k}\uparrow}
%c^{\dag}_{-\mathbf{k}\downarrow}\smash{+}\mathrm{H.c.}\right]\nonumber\\
&-&\!\!\sum_{\mathbf{k,k^{\prime}}} V_{\mathbf{k}\mathbf{k}^{\prime}}
<c^{\dag}_{\mathbf{k}\uparrow}c^{\dag}_{\mathbf{-k}\downarrow}> 
<c_{\mathbf{-k}^{\prime}\downarrow}c_{\mathbf{k}^{\prime}\uparrow}>
%&-&\!\!\sum_{\mathbf{k}}\left|\Delta(\mathbf{k})\right|^2
%/V_{\mathbf{k}\mathbf{k}}
\end{eqnarray}
To obtain the ground state energy of the mean-field Hamiltonian
(\ref{Ham-MF}) we use the equation-of-motion method \cite{Bruus,LQ06} for
the imaginary-time ($\tau=it$) normal ($G_{\sigma\sigma^{\prime}}$)
and anomalous ($F_{\sigma\sigma^{\prime}}$) Green's functions, defined in the usual way as\cite{Bruus} as 
time-ordered expectation values of the fermion creation and annihilation operators:
$G_{\sigma,\sigma^{\prime}} \equiv - \langle T_{\tau}c_{{\bf k}\sigma}(\tau) c^{\dag}_{{\bf k}\sigma^{\prime}(0)} \rangle$, and 
$F_{\sigma,\sigma^{\prime}} \equiv - \langle T_{\tau}c^{\dag}_{{\bf -k}\sigma^{\prime}}(\tau) c^{\dag}_{{\bf k}\sigma} (0)\rangle$. The equations of motions that follow from these are given by:

\begin{eqnarray}
\partial_{\tau}G_{\sigma\sigma^{\prime}}(\mathbf{k},\tau)&=&\smash{-}\delta(\tau)\delta_{\sigma\sigma^{\prime}}
\smash{-}\xi_{\mathbf{k}\sigma}G_{\sigma\sigma^{\prime}}(\mathbf{k},\tau)\nonumber\\
&+&\Delta_{-\sigma\sigma}(\mathbf{k})
F_{-\sigma\sigma^{\prime}}(\mathbf{k},\tau),\label{G-normal}\\
\partial_{\tau}F_{\sigma\sigma^{\prime}}(\mathbf{k},\tau)&=&
\xi_{\smash{-}\mathbf{k}\sigma}
F_{\sigma\sigma^{\prime}}(\mathbf{k},\tau) \nonumber\\
&+&\Delta^{*}_{\sigma-\sigma}(\mathbf{k})
G_{-\sigma\sigma^{\prime}}(\mathbf{k},\tau).\label{G-anomal}
\end{eqnarray}

Eqs.(\ref{G-normal}) and (\ref{G-anomal}) are Fourier transformed
with $\tau\smash{\rightarrow}i\omega_{n}\smash{\equiv}\nu$ and
$\partial_{\tau}\smash{\rightarrow}\smash{-}\nu$, where
$i\omega_{n}\smash{=}(2n\smash{+}1)\pi T$ are Matsubara frequencies.
This gives:\\
\begin{eqnarray} 
G_{\sigma\sigma}(\mathbf{k},\nu) &=& \frac{- (\xi_{-\mathbf{k}-\sigma}\smash{+}\nu)}
{(\xi_{\mathbf{k}\sigma}\smash{-}\nu)(\xi_{-\mathbf{k}-\sigma}\smash{+}\nu)
+ |\Delta_{\sigma,\smash{-}\sigma}(\mathbf{k})|^2}\label{G-Four}\\
F_{\sigma-\sigma}(\mathbf{k},\nu) &=& \frac{\Delta^{*}_{\sigma-\sigma}(\mathbf{k})} {(\xi_{\mathbf{k}-\sigma}\smash{-}\nu)(\xi_{-\mathbf{k}\sigma}\smash{+}\nu) + |\Delta_{\sigma-\sigma}(\mathbf{k})|^2} \nonumber\\
&=& - F_{-\sigma\sigma}(\mathbf{k},\nu)\label{F-Four}
\end{eqnarray}

In terms of radial and angular parts, the pairing interaction can be written in the form:
\begin{equation}
V_{\mathbf{k}\mathbf{k}^{\prime}} = (4\pi/3)\sum_{m}V_{kk^{'}}Y_{1,m}(\hat{\mathbf{k}})
Y_{1,m}^{*}(\hat{\mathbf{k}^{\prime}})\;\; ,
\end{equation}
with $\hat{\mathbf{k}}$=$(\theta,\phi)$. 
%$V_{kk^{'}}$ a separable form convenient for calculations:
%$V^{\sigma-\sigma}_{kk^{'}}$
%$V_{kk^{'}}=\lambda w(k)w(k^{\prime})$. 
Using $\langle c_{-\mathbf{k}^{\prime}\downarrow}c_{\mathbf{k}^{\prime}\uparrow}\rangle\equiv
F^{*}_{\downarrow\uparrow}(\mathbf{k}^{\prime},\tau\smash{=}0^{-})$
and Eq.(\ref{F-Four}), we obtain, 
$\Delta(\mathbf{k})$=$-(1/k_BT)\sum_{\nu\mathbf{k}^{\prime}}
V_{\mathbf{k}\mathbf{k}^{\prime}}
F^{*}_{\downarrow\uparrow}(\mathbf{k}^{\prime},\nu)e^{\nu 0^{+}}$.
We take $V_{kk^{'}}=\lambda w(k)w(k^{\prime})$, a separable form chosen
for convenience;~\cite{BOT05,TIN05,NOZ85} this does not qualitatively change the physics.
This gives for the gap function:
$\Delta(\mathbf{k})=\sum_{m}\Delta_{m}
w(k)Y_{1m}(\hat{\mathbf{k}})\equiv w(k)\Delta(\hat{k})$, where
$\Delta_{m}=-(1/k_BT)\sum_{\nu \mathbf{k}^{\prime}}\lambda
w(k^{\prime}) Y^{*}_{1m}(\hat{\mathbf{k}^{\prime}})
F^{*}_{\downarrow\uparrow}(k^{\prime},\nu)e^{\nu 0^{+}}$. Using the Fourier transform
$F_{\downarrow\uparrow}(k,\nu)$ from Eq.(\ref{F-Four}), and the definition of $G_{\sigma\sigma}$ above,  we deduce equations for the $\ell=1;m=0,\pm 1$ gap amplitudes,
%\begin{widetext}
\begin{equation}\label{gapeqn}
\Delta_{m}=-\!\lambda\!\sum_{\mathbf{k}}\! w(k)Y^{*}_{1m}
(\hat{\mathbf{k}})\Delta(\mathbf{k})[n_{F}(E^{-}_{k})\smash{-}n_{F}(E^{+}_{k})]/(2E_{k})
\end{equation}
%\end{widetext}
and for the particle number densities,
\begin{equation}\label {numbereq}
n_{\sigma}=\sum_{k}\left<c_{k\sigma}^+c_{k\sigma}\right>
              =\sum_{k}G_{\sigma\sigma}(k,\tau=0^-).
\end{equation}
Above, $E_{k}=[{\xi_{k}^{2}+|\Delta^{2}(\mathbf{k})}|]^{1/2}$, with
$\xi_{k}$=$(\xi_{k\uparrow}\smash{+}\xi_{k\downarrow})/2$=$\epsilon_{k}-\mu$
and $\mu$=$(\mu_{\uparrow}\smash{+}\mu_{\downarrow})/2$;
$n_F(E_k^{\pm})$ are Fermi functions, with $E_k^{\pm} = -h \pm E_k$, where $h=(\xi_{k\downarrow}-\xi_{k\uparrow})/2=(\mu_{\uparrow}-\mu_{\downarrow})/2$.
At $T$=$0$ and for $h$$>$$0$, the expectation value of the grand canonical potential
is given by:
\begin{eqnarray}
E_{g}&\equiv&\langle{\mathcal{H}}\rangle\!=\!\sum_{k}
\left\{\xi_{k}\smash{-}E_{k}
\smash{+}\frac{|\Delta(\mathbf{k})|^{2}}{2\epsilon_{k}}\right\}\nonumber\\
&+&\!\!\sum_{k}\left\{(E_{k}\smash{-}h)\theta(\smash{-}E_{k}\smash{+}h)\right\}
-\frac{1}{g}\sum_{m}|\Delta_{m}|^{2}\label{GC}
\end{eqnarray}
Following standard practice, the coupling $\lambda$ has been expressed in terms 
of the ``regularized'' interaction $g$: $1/g$=$1/\lambda$+$(1/(2\pi\hbar)^{3})\int w^2(k)
d^3\mathbf{k}/2\epsilon_{k}~$\cite{ISHK06}. For $w(k)$,
we adopt the N-SR form~\cite{NOZ85}:
$w(k)$=$k_{0}k/(k^{2}_{0}+k^{2})$.

\section{Free Energy Analysis of Ground State} 

\subsection{Free Energy to Quartic Order}

First, we carry out an {\it analytic study} of the p-wave superfluid ground state structure by constructing a
free energy  to quartic order.
We assume that $0\le|\Delta(\mathbf{k})|\ll\xi_{k}$, and expand $E_k$ in Eq.(\ref{GC}) in powers of
$|\Delta(\mathbf{k})|$ keeping terms to 4th order: $E_{k}\sim|\xi_k|
[1\smash{+}|\Delta(\mathbf{k})|^{2}/2|\xi_k|^{2}\smash{-}|\Delta(\mathbf{k})|^{4}/8|\xi_k|^{4}]$.
Note that since $\xi_k \gg |\Delta(\mathbf{k})|$,  $\xi_k = 0  \;(\epsilon_k = \mu)$ is excluded
from the expansion. 
The reasoning is similar to that in Ref. 11, and can be seen on noting the following: $\sum_{E_k>h} E_k=\sum_{E_k>h} \sqrt{|\Delta_k|^2+\xi_k^2}=\sum_{|\Delta_k|\backsim
     |\xi_k|}\sqrt{|\Delta_k|^2+|\xi_k|^2}+\sum_{|\Delta_k|\ll|\xi_k|}\sqrt{|\Delta_k|^2+\xi_k^2}
     =\sum_{|\Delta_k|\backsim |\xi_k|,|\xi_k|>h}|O(\Delta_k|)+\sum_{|\Delta_k|\ll|\xi_k|,\xi_k>h}\sqrt{|\Delta_k|^2+\xi_k^2}$. Owing to negligible integration domain, and small integrand, the first term is negligible compared to the second, giving
     $\sum_{E_k>h} E_k\simeq\sum_{|\Delta_k|\ll|\xi_k|,\xi_k>h}\sqrt{|\Delta_k|^2+\xi_k^2}$; 
      $\xi_k = 0$ is thus excluded.
Substituting this in Eq.(\ref{GC}) and
converting momentum sums to integrals we obtain:

\begin{equation}
E_{g}=\alpha\int|\Delta(\hat{k})|^{2}d\Omega+\beta\int|\Delta(\hat{k})|^{4}d\Omega+\gamma,
\end{equation}
where
\begin{eqnarray}
\alpha&=&(1/(2\pi\hbar)^{3})\int_{E_{k}<h}w^{2}(k)k^{2}dk/2\epsilon_{k}\nonumber\\
&+&(1/(2\pi\hbar)^{3})\int_{E_{k}>h}
w^{2}(k)k^{2}dk(1/2\epsilon_{k}\smash{-}1/2|\xi_{k}|)-1/g\nonumber\\
\beta&=&(1/(2\pi\hbar)^{3})\int_{E_{k}>h}
w^{4}(k)k^{2}dk/8|\xi_{k}|^{3}\\
\gamma&=&(4\pi/(2\pi\hbar)^{3})\int_{E_{k}<h}(\xi_{k}\smash{-}h)k^{2}dk\nonumber.
\end{eqnarray}

Existence of a superfluid phase requires
$\alpha$$<$$0$, otherwise minimization of $E_{g}$ forces the gap
to vanish. The polarization ($P=(n_\uparrow-n_\downarrow)/(n_\uparrow + n_\downarrow)$)
dependence of $E_{g}$ is contained in $\alpha, \beta, \gamma$, which
depend on $h, \mu$ and $w(k)$; $\alpha$ alone depends explicitly on
the coupling $g$. Accordingly, the $\Delta_m$'s are sensitive to changes in $P$.

For fixed $\mu$, $h$, and keeping $E_k$ to $\mathcal{O}(\Delta(\mathbf{k})^4)$, 
we obtain an analytic expression for the ground state energy, $E_{G}$, in
terms of the pairing amplitudes $\Delta_m$:
\begin{eqnarray}\label{simple}
E_{G}=-\frac{\alpha^{2}}{8\beta}+ \gamma +
2\beta\left(|\Delta_{0}|^{2}+|\Delta_{1}|^{2}
+|\Delta_{-1}|^{2}+\frac{\alpha}{4\beta}\right)^{2}\nonumber\\
+\beta(|\Delta_{0}|^{2}\smash{-}2|\Delta_{1}||\Delta_{-1}|)^{2}
\smash{+}4\beta(1\smash{-}t)|\Delta_{0}|^{2}|\Delta_{1}||\Delta_{-1}|
\end{eqnarray}
where $t$=$\cos{\theta}$, with  $\theta =
\sphericalangle (\Delta_{0}\Delta^{*}_{1},\Delta^{*}_{0}\Delta_{-1}) = $ 
%(\widehat{\Delta_{0}\Delta^{*}_{1},\Delta^{*}_{0}\Delta_{-1})}$ =
$2\phi_{0}\smash{-}\phi_{1}\smash{-}\phi_{-1}$; $\phi_m$'s
are the phases associated with the gap amplitudes $\Delta_m$. 

\subsection{Energy Minimization: Global and Local Minima}

Minimizing $E_G$ in (\ref{simple}) with respect to $\Delta_m$'s, we find that for a stable p-wave superfluid phase to exist, the following conditions have to be
satisfied simultaneously 
\begin{eqnarray}\label{conditions}
(a)&|\Delta_{0}|^{2}\smash{+}|\Delta_{1}|^{2}\smash{+}|\Delta_{-1}|^{2} &=\smash{-}\alpha/4\beta  \equiv R^2  \;\;\;(\rm sphere)\nonumber\\
(b)&|\Delta_{0}|^{2}\smash{-}2|\Delta_{1}||\Delta_{-1}| &= 0 \;\;\;(\rm plane)\\
(c)&(1\smash{-}t) |\Delta_{0}|^{2}|\Delta_{1}||\Delta_{-1}| &= 0\nonumber
\end{eqnarray}

For all three $\Delta_m$'s non-zero, conditions 14(a) and 14(b) give a semicircle formed by the
intersection of the surface of a sphere of radius $R$, ($R^{2}=-\alpha/4\beta$), in the space
of $\Delta_{1}, \Delta_{-1}, \Delta_{0}$, 
with a plane defined by $|\Delta_{1}|+|\Delta_{-1}|$=$R$. The points spanning the semicircle represent a multitude of "mixed" superfluid states of the form $a_0\Delta_0 +a_1\Delta_1 + a_{-1}\Delta_{-1}$; $a_0, a_1, a_{-1}$ being constants.  A 3D geometric representation of the states is shown in Fig.1. The third condition 14(c)
imposes the constraint, $t\equiv \cos\theta =1$, i.e. the relative phase
$\theta = (\phi_{0}\smash{-}\phi_{1})\smash{-}(\phi_{-1}\smash{-}\phi_{0})= 2n\pi$.
$t$=$1$ corresponds to a parallel orientation of vectors
$\Delta_{0}\Delta^{*}_{1}$ and $\Delta^{*}_{0}\Delta_{-1}$.
The calculated ground state (GS) {\it global minimum} energy,
\begin{equation}\label{global}
E^{gl}_G = -\alpha^{2}/8\beta + \gamma,  \;\;\;\;(\alpha < 0),
\end{equation}
is completely determined by $\alpha$, $\beta$, and $\gamma$ given earlier. 

The phase condition 14(c)  is always satisfied for the states, $\Delta_{\pm 1}$, at the endpoints A and B of the semicircle.
%and so the phase condition is automatically satisfied.
But, the energies of the states on the semicircle, other than A, B, oscillate with $\theta$, and attains the global minimum $E^{gl}_G$ only for $t=1$, i.e. $\theta=2n\pi$ (Fig. 3a). The  oscillation amplitude depends on the specific state; e.g. state C at the top of the projected figure (Fig.1) has the maximum amplitude. Representative states A,B,C on semicircle (Fig.1) are given by: (A) $|\Delta_{0}|\smash{=}|\Delta_{-1}|$=$0$ and
$|\Delta_{1}|^2$=$-\alpha/4\beta$; (B) $|\Delta_{0}|\smash{=}|\Delta_{1}|\smash{=}0$ and
$|\Delta_{-1}|^{2}$=$-\alpha/4\beta$; (C) $|\Delta_{0}|^{2}$=$-\alpha/8\beta$ and
$|\Delta_{1}|^{2}\smash{=}|\Delta_{-1}|^{2}\smash{=}-\alpha/16\beta$.

\begin{figure}[t]
{\scalebox{0.65}{\includegraphics[clip,angle=0]{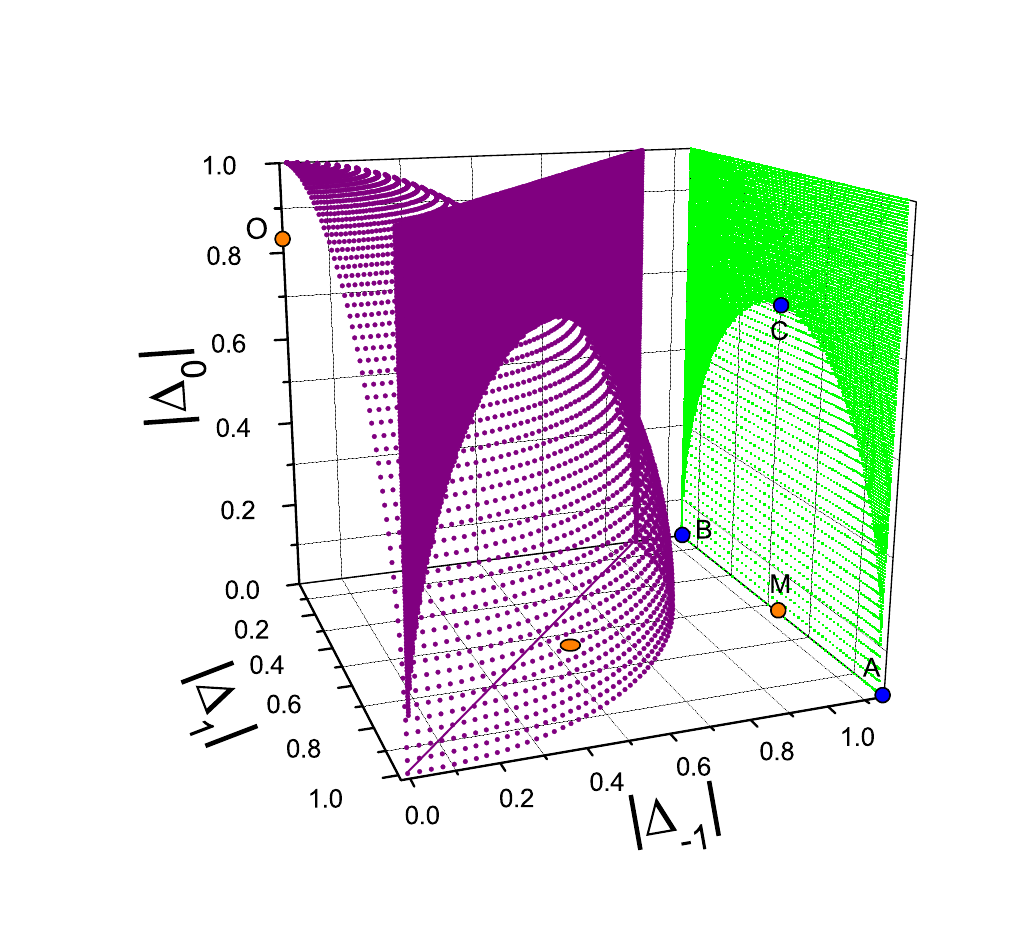}}}
\caption{(Color online) Geometric representation of p-wave
states in population-imbalanced Fermi system. The semicircle formed by the intersection of sphere surface with plane represent a continuum of states exhibiting global energy minimum (see text). For clarity, plane containing semicircle is shown separately (green). O (on $\Delta_0$ axis) and M (in $\Delta_1$-$\Delta_{-1}$ plane) are states with local energy minimum. $\Delta_m (m=0,\pm 1)$'s are normalized to the sphere radius $\sqrt{- \alpha/4\beta}$; $\alpha < 0$.}
\label{Fig1}
\end{figure}

%Conditions (a) and (b) together represent a semicircle formed by the
%intersection of the surface of a sphere of radius $R$, $R^{2}=-\alpha/4\beta$,
%with a plane defined by $|\Delta_{1}|+|\Delta_{-1}|$=$R$. All points on the semicircle
%must necessarily satisfy the phase constraint (c). They represent a multitude of
%``mixed'' superfluid states of the form $a\Delta_0 +b\Delta_1 +c\Delta_{-1}$.
%A 3D geometric representation of the states is shown in Fig. 1.
%The GS global energy minimum condition is satisfied only for $t=1$ in Eq. 9c, i.e. when the %relative phase
%$\theta = (\phi_{0}\smash{-}\phi_{1})\smash{-}(\phi_{-1}\smash{-}\phi_{0})= 2n\pi$.
%For the states
%$\Delta_{\pm 1}$ (points  A and B on the semicircle in Fig 1), this is always satisfied.
%For the other states on this curve, containing varying admixtures of $\Delta_m$'s,
%$E_G$ fluctuates with $\theta$, and attains the global minimum
%only for $\theta=2n\pi$ (Fig. 3a); the fluctuation amplitude
%depends on the specific state chosen; e.g. C at the top of the projected figure
%has the maximum amplitude variation. 

We obtain {\it other} minimum energy solutions by considering special cases in Eq. \ref{simple}.
(i) $|\Delta_1| = 0 = |\Delta_{-1}|$ and $|\Delta_0| \ne 0$: we find a unique solution, $|\Delta_0|^2 = - \alpha/6\beta$, shown as point O in Fig.1. (ii) $|\Delta_0| = 0$: this leads to equations for two ellipses in the $\Delta_1 - \Delta_{-1}$ plane:
\begin{eqnarray}
|\Delta_1|^2 + 2 |\Delta_{-1}|^2 + \frac{\alpha}{4\beta} = 0  \;\;\;\; (\rm {ellipse 1})\nonumber\\
2 |\Delta_1|^2 + |\Delta_{-1}|^2 + \frac{\alpha}{4\beta} = 0  \;\;\;\; \rm{(ellipse 2})
\end{eqnarray}
The intersection of the two ellipses gives the solution: $|\Delta_1|^2 = |\Delta_{-1}|^2 =
- \alpha/12\beta$ and $|\Delta_0| = 0$; this is shown as point M in Fig.1. Like A and B, 
states O and M automatically satisfy phase constraint (Eq. 14(c)).
We find these to exhaust all possible independent solutions. Incidentally, 
states A (B)  respectively lie at the intersection of ellipse 1 (ellipse 2), and the sphere and  plane of Fig.1. The ground state energies for states O and M are found to equal each other, and lie higher than the global minimum $E^{gl}_G$; we denote this as {\it local} minimum energy: 
\begin{equation}
E_G^{loc} = - \frac{\alpha^2}{12 \beta} + \gamma \;\;\;\; (\alpha < 0)
\end{equation}

We checked the stability of the states, by examining  the stability 
matrix or Hessian, H: $(\partial^2{E_G}/\partial{\Delta_{m_i}}\partial{\Delta_{m_j}})$. We find that
$\rm{det} [H] = 0$ for states A and B, and $\rm{det} [H] > 0$ for states O and M.
In the language of Catastrophe Theory~\cite{catas}, A, B can be identified as non-Morse "critical" points, and O, M as Morse "critical" points. This is based on Morse lemma for Morse critical points, and Thom's theorem and splitting lemma for non-Morse critical points.
det [H] = 0 occurs when one or more of the eigenvalues of the stability Hessian matrix H is zero. In this case,
Thom's splitting lemma can be used to represent the stability matrix H in a block-diagonal form, wherein H 
is split into blocks with non-zero  determinants.~\cite{catas}

%~\cite{critical}.

As can be seen from the expression for $\gamma$, and the definition of $\mu$ and $h$,
in the limit of zero polarization, $P \rightarrow 0$, the GS energy
coefficient $\gamma \rightarrow 0$, while $\alpha<0,  \beta>0$ remain
finite, but with different numerical values. Accordingly, the expressions for $E^{gl}_G$ and $E_G^{loc}$ remain essentially the same for $P \rightarrow 0$, but attain numerical values different from that for $P\ne0$. The states A and B
(equivalent to A by symmetry) correspond to the
finding of Ref.~\cite{TIN05} in $P=0$ case. 
%As discussed, our work reveals additional superfluid states and features.
%set of ``mixed'' superfluid states with global minimum energy, as well those with local %minimum energy. 

\section{Self-consistent Solutions of Coupled Gap and Number Equations} 

Next, we self-consistently  solve the set of three coupled gap equations (Eq. (8), and the two
number equations (Eq. (9)). This is done for for fixed
population imbalance $P=(n_\uparrow-n_\downarrow)/(n_\uparrow + n_\downarrow)$, and p-wave coupling parameter $g$.
($g$ is related to the triplet scattering parameter: $g=25\/k_F^3a_t/8\pi$ 
for a cutoff $k_0=10k_F$). These detailed solutions of the p-wave superfluid states confirm the results  obtained using free energy considerations (Sec.  III above), and also reveal additional interesting features.
%We solve self-consistently the three gap equations (Eq. (5))  and two number equations (Eq. (6)), for fixed
%population imbalance $P=(n_\uparrow-n_\downarrow)/(n_\uparrow + n_\downarrow)$, and p-wave coupling %parameter $g$.
%($g$ is related to the triplet scattering parameter: $g=25\/k_F^3a_t/8\pi$  for a cutoff $k_0=10k_F$).
We obtain the gap amplitudes $\Delta_m$, and the chemical potentials $\mu_{\sigma}$ for different
$P$ and $1/k_F^3 a_t$.
Using these, we obtain values of the ground state energy $E_g$ from Eq. (7), as well as those of 
the coefficients $\alpha,\beta,\gamma$ that determine $E_g$. 
The agreement between our self-consistent solutions and analytical results for the ground state energies  
is good. We note that fixed ($P, n = n_\uparrow + n_\downarrow$) is equivalent to
fixed ($\mu, h$) used in analytic study above.

\subsection{Phase Stability and Construction of Phase Diagram}

To check for {\it stability} of the p-wave states obtained from the self-consistent solutions, we enforce
that the stability matrix $(\partial^2{E_G}/\partial{\Delta_{m_i}}\partial{\Delta_{m_j}})$
is positive definite;  and that $\delta p/ \delta h >0$.
Based on this, we construct a polarization ($P$) - coupling ($1/k_F^3 a_t$) {\it phase diagram} in BEC-BCS crossover regime (Fig.2).
Instability in the stability matrix can just indicate instability to 
another phase, but not the nature of the phase. However, this is sufficient to map out the phase boundary, e.g.
the critical polarization line $P_c$, where superfluid gap vanishes. Above $P_c$, the system is in the normal phase. Below $P_c$, the stability of phases SF1 and SF2 are carefully examined in turn in the coupling-polarization space; this allows us to map out the full phase diagram.

In Fig. 2, SF1 denotes the stable superfluid phase corresponding to the states on
the `semicircle' in Fig.1 that give GS {\it global} minimum, $E^{gl}_G$ (Eq. 15),
 i.e. with relative phase $\theta$=$2n\pi$ among the gap parameters. $E^{gl}_G$ and $E^{loc}_G$ (Eq. 17) are given by the polarization-dependent parameters,  $\alpha, \beta, \gamma$.
With increased polarization, $\alpha, \beta, \gamma$ (given by Eqs. 12 above), that determine  $E^{gl}_G$ and $E^{loc}_G$, change so that 
the superfluid phase SF2, corresponding to states with {\it local} minimum, attain energy lower than that of  SF1, thereby becoming stable. This suggests the interesting possibility that at T=0, polarization may drive a quantum phase transition from SF1 to SF2.  It may also be possible to access SF2 at T$\ne$0.
%SF1 becomes unstable, and gives way to the superfluid phase
%SF2, corresponding to states with the {\it local} minimum discussed above.
%In Fig. 2, states $\Delta_1$ and $\Delta_0$ were chosen in
%SF1 and SF2 respectively; other choices of allowed global and local minimum states
%give qualitatively the same phase diagram. 
At even larger polarizations, for the same reason,  
SF2 becomes unstable to phase separation (PS).  SF1, SF2, and PS occupy a relatively much
narrower part of the phase diagram on the BCS side compared to the BEC side.
In our two-component system with inter-species interaction, PS persists into
full polarization, P=1.
This is reasonable because at P=1, the system is essentially a one-component
system in which the absence of minority species atoms makes inter-species interaction
inoperative. Such a system can exhibit superfluidity only under the effect of
intra-species interactions.

We note that since we have considered p-wave ($\ell$ = 1) pairing with unlike spin components 
($S=1$, $m_s = 0$),  the gap functions can be classified as $\ell$ = 1 representations of the O(3)
group embedded in O(6). Eq. (13) above reflects this O(3) symmetry. Accordingly,
the states in SF1, in particular the $\Delta_{1, \pm1}$  states A and B, can be identified with the known "axial" or "ferromagnetic" phase, while the states in SF2, such O and M,  with the known "polar" phase.

\begin{figure}[t]
{\scalebox{0.65}{\includegraphics[clip,angle=0]{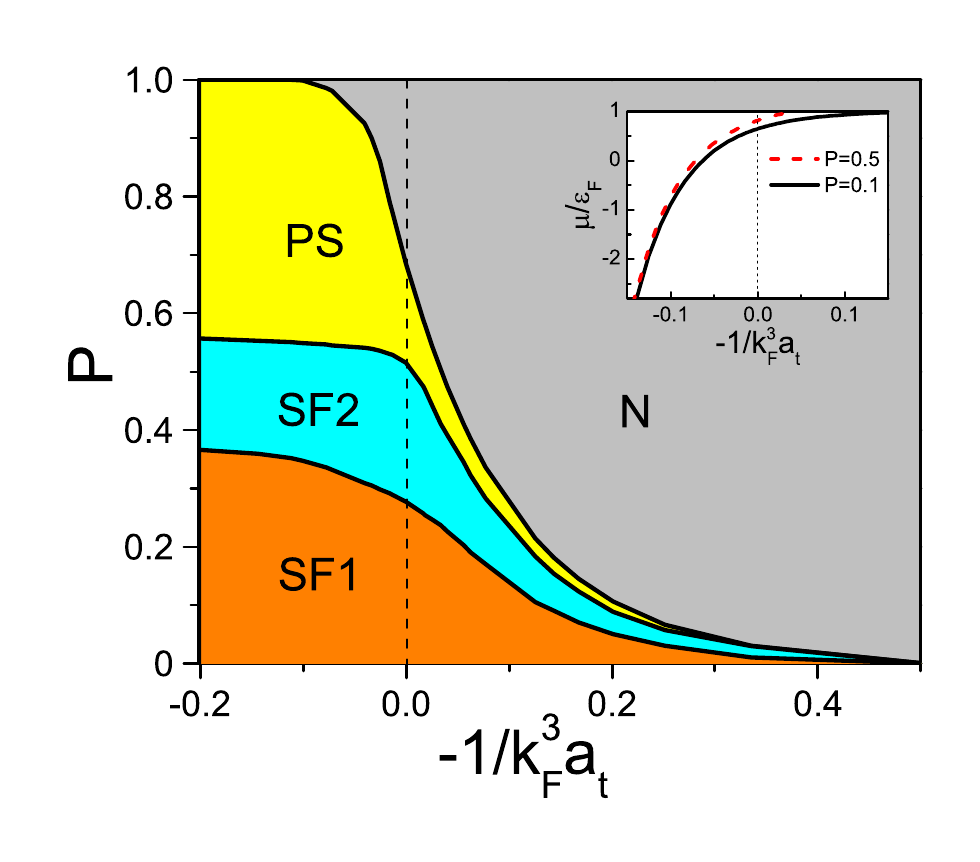}}}
\caption{(Color online) Polarization $P$ vs p-wave coupling $-1/(k^{3}_{F}a_{t})$
phase diagram of two-component Fermi system with $p$-wave pairing.
Shown are normal (N),  p-wave superfluid phases (SF1; SF2), and phase separation (PS).
Unitarity limit is shown by the dashed vertical line. Inset:
Calculated chemical potential vs coupling across BCS and BEC regimes
for $P$ = 0.1 (solid line), 0.5 (dashed line). The Fermi energy, $\epsilon_F$, 
is given by the line $\mu/\epsilon_F = 1$.} \label{Fig2}
\end{figure}

The inset in  Fig. 2 shows the calculated behavior of
chemical potential $\mu$ across the  BEC-BCS regime.
It  deviates significantly from the Fermi energy (given by $\mu/\epsilon_F$ = 1 line)
in a wider region around the BEC-BCS crossover, even on the BCS side, and drops much more rapidly to negative values on the BEC side compared to the s-wave case.
For sufficiently weak coupling in the BCS regime, $\mu$ approaches
Fermi energy $\epsilon_{F}$.

\begin{figure}[t]
{\scalebox{0.65}{\includegraphics[clip,angle=0]{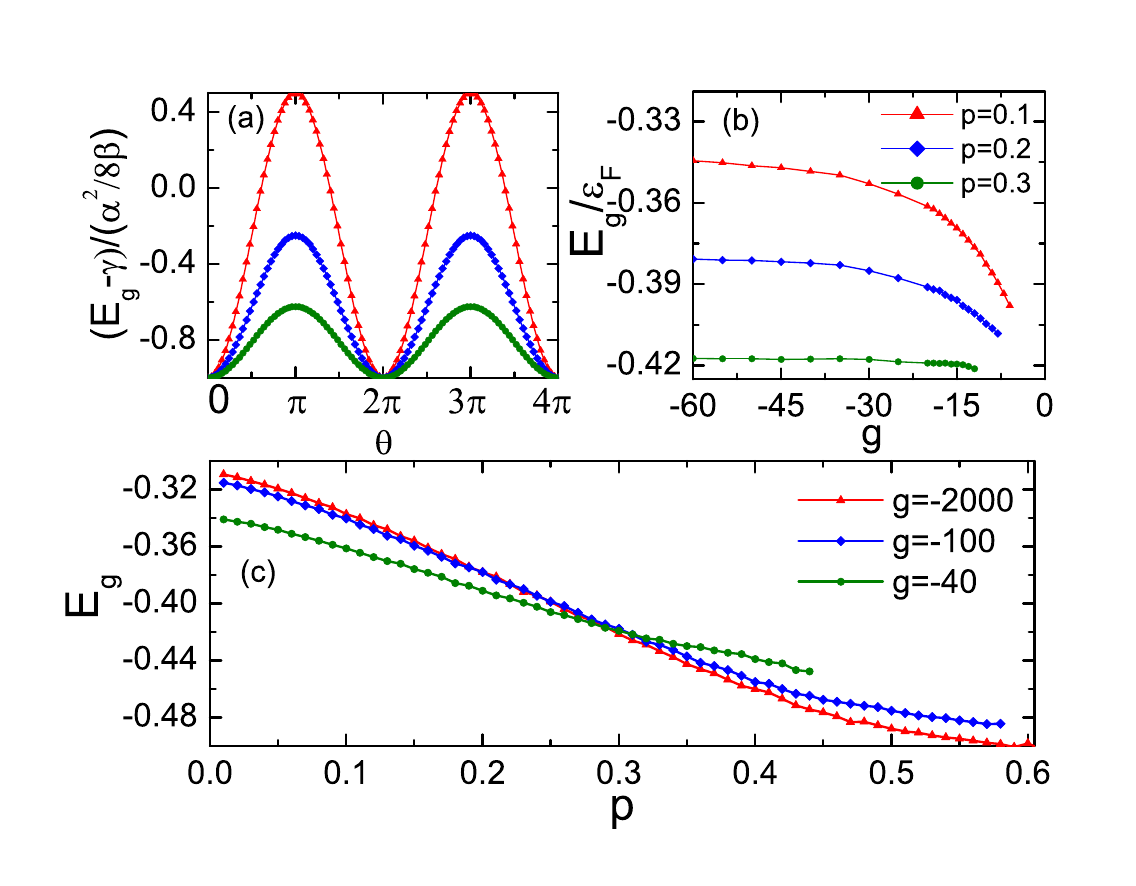}}}
\caption{(Color online) (a) Calculated ground state energy (scaled to global minimum) vs.
relative phase angle $\theta$ (see text).
%between gap parameters for different
%$Y_{11}$-like pairing wavefunctions.
Curve with maximal amplitude corresponds to the state C in Fig 1: $|\Delta_{0}|^{2}$=$-\alpha/8\beta$,
$|\Delta_{1}|^{2}$=$|\Delta_{-1}|^{2}$=$-\alpha/16\beta$. Other two curves respectively correspond to other representative states on the arc in
Fig. 1: $|\Delta_{0}|^{2}$=$-\alpha/11.3\beta$,
$|\Delta_{1}|^{2}$=$-\alpha/76.3\beta$,
$|\Delta_{-1}|^{2}$=$-\alpha/6.73\beta$, and
$|\Delta_{0}|^{2}$=$-\alpha/16\beta$,
$|\Delta_{1}|^{2}$=$-\alpha/186\beta$,
$|\Delta_{-1}|^{2}$=$-\alpha/5.49\beta$. (b) Calculated
ground state energy $E_g$ vs. coupling $g$ for different polarizations $P$.
(c) Calculated $E_g$ vs.  $P$ for
different  $g$. } \label{Fig3}
\end{figure}

\subsection{Variation of Ground state energy with Phase angle, Coupling and Polarization}

Except for A, B, the continuum of ``mixed'' SF states on the semicircle (Fig.1) are characterized by a relative phase $2n\pi \le \theta \le (2n+2)\pi$, so
that the energies of these states are expected to oscillate with $\theta$. Results for three representative cases, obtained  from our numerical calculations, are shown in Fig 3(a). 
For $\theta \ne 2n\pi$, the energies lie higher than the GS global minimum; the maximum amplitude occurring for the state C in Fig.1. This raises the possibility of  
observing  $\theta$-oscillation in each of the states on the semicircle using phase sensitive experimental technique(s). They could also be accessed at finite-T.
%accessing the multitude of $\theta$-dependent states, corresponding to each of the states on the semicircle, using phase sensitive experimental techniques. 

Figs. 3b, 3c respectively show the variation of $E_{g}$ with coupling $g$
at fixed polarizations $P$, and with $P$ for
fixed $g$. $E_g$ is normalized to the Fermi energy $\epsilon_F$ =
$\hbar^2 k_F^2/2m$, with $2k_F^3=k_{F\uparrow}^3+k_{F\downarrow}^3$.
For a given $P$, $E_{g}$ becomes less negative as $g$ approaches
unitarity; the trend is more noticeable for smaller $P$'s. The energy curves in Fig. 3(b) 
would terminate at some non-zero coupling, and hence not expected to cross.
Extrapolated to normal regime, energies corresponding to different polarizations 
would be different, and also not cross.
For a given $g$, $E_g$ lies higher for smaller $P$, presumably
due to the lower majority-species band becoming progressively more
occupied with increasing $P$, thereby lowering $E_{g}$ with increasing polarization.
For small $P$, $E_g$ becomes less negative with increasing $g$. This trend is reversed
for $P \ge 0.3$. The crossing at $P \approx 0.3$ is suggestive of a possible scaling
using stretches in $P$ and $E_g$.

\section{Summary} 

We have presented several new results for p-wave pairing in two-component population imbalanced Fermi systems for the case when inter-species pairing interaction is dominant, across the entire BCS-BEC regimes. The  ground state structure 
as a function of population imbalance is rich, involving various sub-states of orbital angular momentum $\ell=1$. We find states giving both global and local energy minimum that we associate with non-Morse and Morse critical points. The 3D geometric rendering of our analytic solutions of the SF states provides added insight into these. Our detailed numerical calculations suggest a possible quantum phase transition between two superfluid phases driven by polarization. The energies of a multitude of "mixed" SF states show oscillations with a relative phase 
angle, that may be observed in phase sensitive experiments.
Insight into the nature of the orbital part of our superfluid states may be gained
from measuring the angular dependence of momentum distributions; from molecular spectroscopy using light radiation; or possibly measurements of zero sound attenuation. While this work has not considered
possibilities such as the FFLO states with non-zero center-of-mass momentum (mentioned in Sec. II),
or "breached pair" states~\cite{Liu}, our findings suggest that the problem of $q=0$ unconventional pairing 
for population imbalanced systems, even at the mean-field level, is interesting in itself. Additionally, this  work may form a basis for exploring the possibility of non-s-wave "breached pair" superfluidity in BCS or BEC region. Our work should be of interest to other unequal population Fermi systems, especially  where
p-wave intra-species couplings are small or negligible.

\section{Acknowledgements}

We acknowledge helpful discussions with Jason Ellis, Randy Hulet, Harry Kojima,
and Adriana Moreo. The work was partly supported by
funding from ICAM. One of us (F. Popescu) acknowledges an ICAM Fellowship.

\suppressfloats
\end{document}